\documentclass[reprint,onecolumn,notitlepage,footinbib,aps,superscriptaddress]{revtex4-1}
\usepackage{biolinum}
\usepackage[latin9]{inputenc}
\usepackage{geometry}
\usepackage{color}
\usepackage{calc}
\usepackage{bbold}
\usepackage{float}
\usepackage{textcomp}
\usepackage{amsmath}
\usepackage{amssymb}
\usepackage{wasysym}
\usepackage{longtable}
\usepackage{stackrel}
\usepackage{lipsum}
\usepackage{multirow}
\usepackage{graphicx}
\usepackage{xcolor}

\usepackage{makecell}
\usepackage{epstopdf}

\PassOptionsToPackage{normalem}{ulem}
\usepackage{ulem}

\begin{document}

\title{Long-distance transmission of quantum key distribution coexisting with classical optical communication over weakly-coupled few-mode fiber}

\author{Bi-Xiao Wang}
\affiliation{Hefei National Laboratory for Physical Sciences at Microscale and Department of Modern Physics, University of Science and Technology of China, Hefei, Anhui 230026, P.~R.~China}
\affiliation{CAS Center for Excellence and Synergetic Innovation Center in Quantum Information and Quantum Physics, University of Science and Technology of China, Hefei, Anhui 230026, P.~R.~China}

\author{Yingqiu Mao}
\affiliation{Hefei National Laboratory for Physical Sciences at Microscale and Department of Modern Physics, University of Science and Technology of China, Hefei, Anhui 230026, P.~R.~China}
\affiliation{CAS Center for Excellence and Synergetic Innovation Center in Quantum Information and Quantum Physics, University of Science and Technology of China, Hefei, Anhui 230026, P.~R.~China}

\author{Lei Shen}
\affiliation{State Key Laboratory of Optical Fiber and Cable Manufacture Technology, Yangtze Optical Fiber and Cable Joint Stock Limited Company, Wuhan, 430070, P.~R.~China}

\author{Lei Zhang}
\affiliation{State Key Laboratory of Optical Fiber and Cable Manufacture Technology, Yangtze Optical Fiber and Cable Joint Stock Limited Company, Wuhan, 430070, P.~R.~China}

\author{Xiao-Bo Lan}
\affiliation{State Key Laboratory of Optical Fiber and Cable Manufacture Technology, Yangtze Optical Fiber and Cable Joint Stock Limited Company, Wuhan, 430070, P.~R.~China}

\author{Dawei Ge}
\affiliation{State Key Laboratory of Advanced Optical Communication Systems and Networks, Peking University, Beijing 100871, P.~R.~China}

\author{Yuyang Gao}
\affiliation{State Key Laboratory of Advanced Optical Communication Systems and Networks, Peking University, Beijing 100871, P.~R.~China}

\author{Juhao Li}
\affiliation{State Key Laboratory of Advanced Optical Communication Systems and Networks, Peking University, Beijing 100871, P.~R.~China}

\author{Yan-Lin Tang}
\affiliation{QuantumCTek Corporation Limited, Hefei, Anhui 230088, P.~R.~China}

\author{Shi-Biao Tang}
\affiliation{QuantumCTek Corporation Limited, Hefei, Anhui 230088, P.~R.~China}

\author{Jun Zhang}
\affiliation{Hefei National Laboratory for Physical Sciences at Microscale and Department of Modern Physics, University of Science and Technology of China, Hefei, Anhui 230026, P.~R.~China}
\affiliation{CAS Center for Excellence and Synergetic Innovation Center in Quantum Information and Quantum Physics, University of Science and Technology of China, Hefei, Anhui 230026, P.~R.~China}

\author{Teng-Yun Chen}
\email{tychen@ustc.edu.cn}
\affiliation{Hefei National Laboratory for Physical Sciences at Microscale and Department of Modern Physics, University of Science and Technology of China, Hefei, Anhui 230026, P.~R.~China}
\affiliation{CAS Center for Excellence and Synergetic Innovation Center in Quantum Information and Quantum Physics, University of Science and Technology of China, Hefei, Anhui 230026, P.~R.~China}

\author{Jian-Wei Pan}
\email{pan@ustc.edu.cn}
\affiliation{Hefei National Laboratory for Physical Sciences at Microscale and Department of Modern Physics, University of Science and Technology of China, Hefei, Anhui 230026, P.~R.~China}
\affiliation{CAS Center for Excellence and Synergetic Innovation Center in Quantum Information and Quantum Physics, University of Science and Technology of China, Hefei, Anhui 230026, P.~R.~China}


\maketitle

Quantum key distribution (QKD) is one of the most practical applications in quantum information processing, which can generate information-theoretical secure keys between remote parties. With the help of the wavelength-division multiplexing technique, QKD has been integrated with the classical optical communication networks. The wavelength-division multiplexing can be further improved by the mode-wavelength dual multiplexing technique with few-mode fiber (FMF), which has additional modal isolation and large effective core area of mode, and particularly is practical in fabrication and splicing technology compared with the multi-core fiber. Here, we present for the first time a QKD implementation coexisting with classical optical communication over weakly-coupled FMF using all-fiber mode-selective couplers. The co-propagation of QKD with one 100 Gbps classical data channel at -2.60 dBm launched power is achieved over 86 km FMF with 1.3 kbps real-time secure key generation. Compared with single-mode fiber, the average Raman noise in FMF is reduced by 86\% at the same fiber-input power. Our work implements an important approach to the integration between QKD and classical optical communication and previews the compatibility of quantum communications with the next-generation mode division multiplexing networks.

\section{Introduction}

Quantum key distribution (QKD) guarantees an information-theoretical secure generation of private keys between distant parties, based on the laws of quantum mechanics\cite{Gisin2002}. In the past 35 years, tremendous achievements of QKD have been accomplished\cite{Scarani2009,Diamanti2016}. Currently, the co-propagation of QKD and classical optical communication in the existing optical fiber-based network infrastructure is one of the most important solutions to promote the industrialization of QKD\cite{Zhang:18}.

The wavelength-division multiplexing (WDM) scheme is a standard technique to combine QKD and classical signals\cite{575910}. In such scheme, the spontaneous Raman scattering (SRS) noise generated by classical optical signals is the dominant impairment for quantum signals\cite{Mao:18}. The SRS noise can be suppressed by three major techniques including simultaneous filtering in the time and frequency domains\cite{doi:10.1063/1.4864398}, increasing the spectral interval between quantum and classical signals\cite{doi:10.1063/1.2117616}, and decreasing the launched power of classical optical communication\cite{Frohlich:17}. 

Recently, space-division multiplexing (SDM) with additional available degrees of freedom of optical photons is attractive to enhance the WDM technique\cite{Xavier2020}. As a primary direction for the future evolution of optical networks, SDM was originally proposed to solve the ``capacity crunch'' of the single-mode fiber (SMF)\cite{Richardson2013}. Currently, SDM based on multi-core fiber for the integration between QKD and classical optical communication has been studied\cite{Dynes:16,Bacco2019,Xavier2020}. 

The more practical approach to implement SDM is using few-mode fiber (FMF), which is also called mode-division multiplexing\cite{Richardson2013}. Compared with multi-core fiber, the manufacturing process of FMF is relatively simple and the existing fusion equipment for SMF can be directly utilized for FMF\cite{8004172,Awaji2018ReviewOS}. Most importantly, the large effective core area of the mode and additional modal isolation of FMF is beneficial to suppress SRS noise\cite{Carpenter:13}.

In this paper, we present for the first time a mode-wavelength dual multiplexing QKD implementation over weakly-coupled FMF coexisting with classical optical communication. Using the all-fiber mode-selective couplers (MSCs)\cite{8535538,Ge2019}, the co-propagation distance of QKD with one 100 Gbps data channel at -2.60 dBm launched power reaches 86 km with a secure key rate of 1.3 kbps. At the same fiber-input power, compared with SMF, SRS noise in FMF is reduced by 86\% in average.

\section{Few-mode fiber characterization}
FMF is based on the orthogonality of different linear-polarized (LP) modes. We use the weakly-coupled three-layer ring-core FMF, which has two circular-symmetric LP modes ($\rm LP_{01}$ and $\rm LP_{02}$) and four degenerate LP modes ($\rm LP_{11}$, $\rm LP_{21}$, $\rm LP_{31}$ and $\rm LP_{12}$)\cite{8535538}. In this experiment, two modes, i.e., $\rm LP_{01}$ and $\rm LP_{02}$, are used due to the effect of degenerate mode. To connect FMF and SMF, fiber-based MSCs are utilized as the mode multiplexer/demultiplexer (MUX/DEMUX), which are fabricated by heating and tapering the FMF and SMF. With the help of MSCs, the mode of optical pulses can be transformed each other between the fundamental mode of SMF and the specific LP mode of FMF\cite{Ge2019}. The attenuation and the insertion loss (IL) of MSC as the MUX/DEMUX in each LP mode are characterized, as listed in Table 1. 

\begin{table}[h!]
\setlength{\abovecaptionskip}{0. cm}
\setlength{\belowcaptionskip}{-0. cm}
\caption{Characteristics of the FMF and mode MUX/DEMUX}
\begin{tabular}{c|c|c|c}
\hline
\multirow{2}{*}{\makecell{Mode type}} &
\multirow{2}{*}{\makecell{\makecell{Attenuation\\ (dB/km)}}}&
\multicolumn{2}{c}{IL of MSC (dB)} \\
\cline{3-4}
  & &mode MUX & mode DEMUX\\
\hline
 $\rm LP_{01}$ & 0.226  & 2.60 & 2.30 \\
\hline
 $\rm LP_{02}$ & 0.257  & 3.70 & 3.20 \\
\hline
\end{tabular}   
\centering         
\end{table}

We first characterize the overall modal isolation provided by both the FMF and MUX/DEMUX between different LP modes. Figure 1 illustrates the schematic structure of the experimental setup and mode patterns observed by a charge-coupled device. A continuous laser at 1546.92 nm of 0 dBm is launched into $a$ or $b$ port, and then, the optical power emitted from $c$ and $d$ ports are measured. The output power difference between $c$ and $d$ ports are calculated and listed in the modal isolation matrix, as given in Table 2.

\begin{figure}[h!]
\centering
\includegraphics[width=1\textwidth]{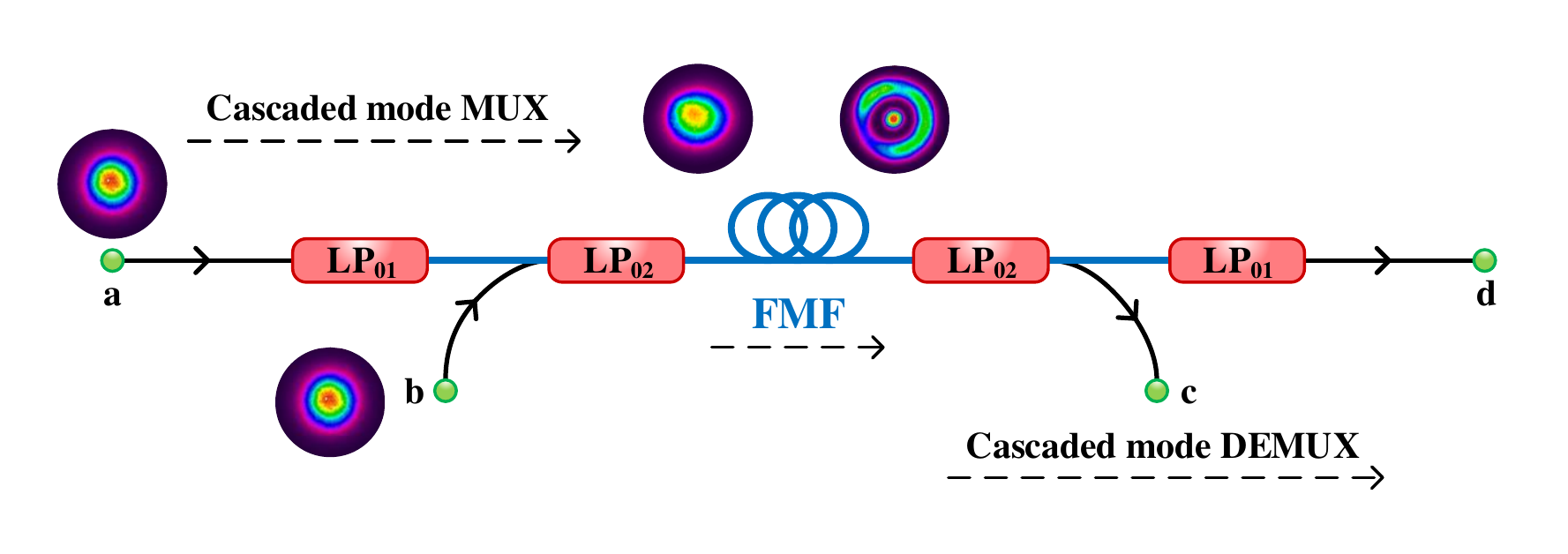}
\caption{Experimental setup for measuring the overall modal isolation. Blue lines indicate FMF and black lines with arrows denote SMF. Patterns at $a$ port and $b$ port correspond to the initial spots of the laser over SMF. The patterns in the middle are the $\rm LP_{01}$ (left) and $\rm LP_{02}$ (right) mode's spots after the mode MUX.}
\end{figure} 

The $\rm LP_{01} in$ ($\rm LP_{02} in$) scheme refers to the case that the classical light launched from $a$ ($b$) port is assigned to the $\rm LP_{01}$ ($\rm LP_{02}$) mode, and the modal isolation from the $\rm LP_{02}$ ($\rm LP_{01}$) mode to the $\rm LP_{01}$ ($\rm LP_{02}$) mode is recorded. The modal isolation of cascaded mode MUX/DEMUX corresponds to the back-to-back (BTB) case. Due to the perturbations of optical fiber such as macrobending and microbending, the overall modal isolation decreases as the transmission distance\cite{doi:10.1002/0471219282.eot158}.

\begin{table}[h!]
\setlength{\abovecaptionskip}{0. cm}
\setlength{\belowcaptionskip}{0. cm}
\caption{Modal isolation of the FMF and mode MUX/DEMUX (dB)}
\begin{tabular}{c|c|c|c|c|c}
\hline
& BTB & 25 km & 50 km & 75 km & 100 km\\
\hline
$\rm LP_{01} in$ & 23.58 & 21.73 & 19.96 & 17.17 & 15.75 \\
\hline
$\rm LP_{02} in$ & 23.20 & 19.25 & 14.75 & 12.59 & 10.35 \\
\hline
\end{tabular}   
\centering         
\end{table}

\section{Experiment}

\begin{figure}[h!]
\centering
\includegraphics[width=1\textwidth]{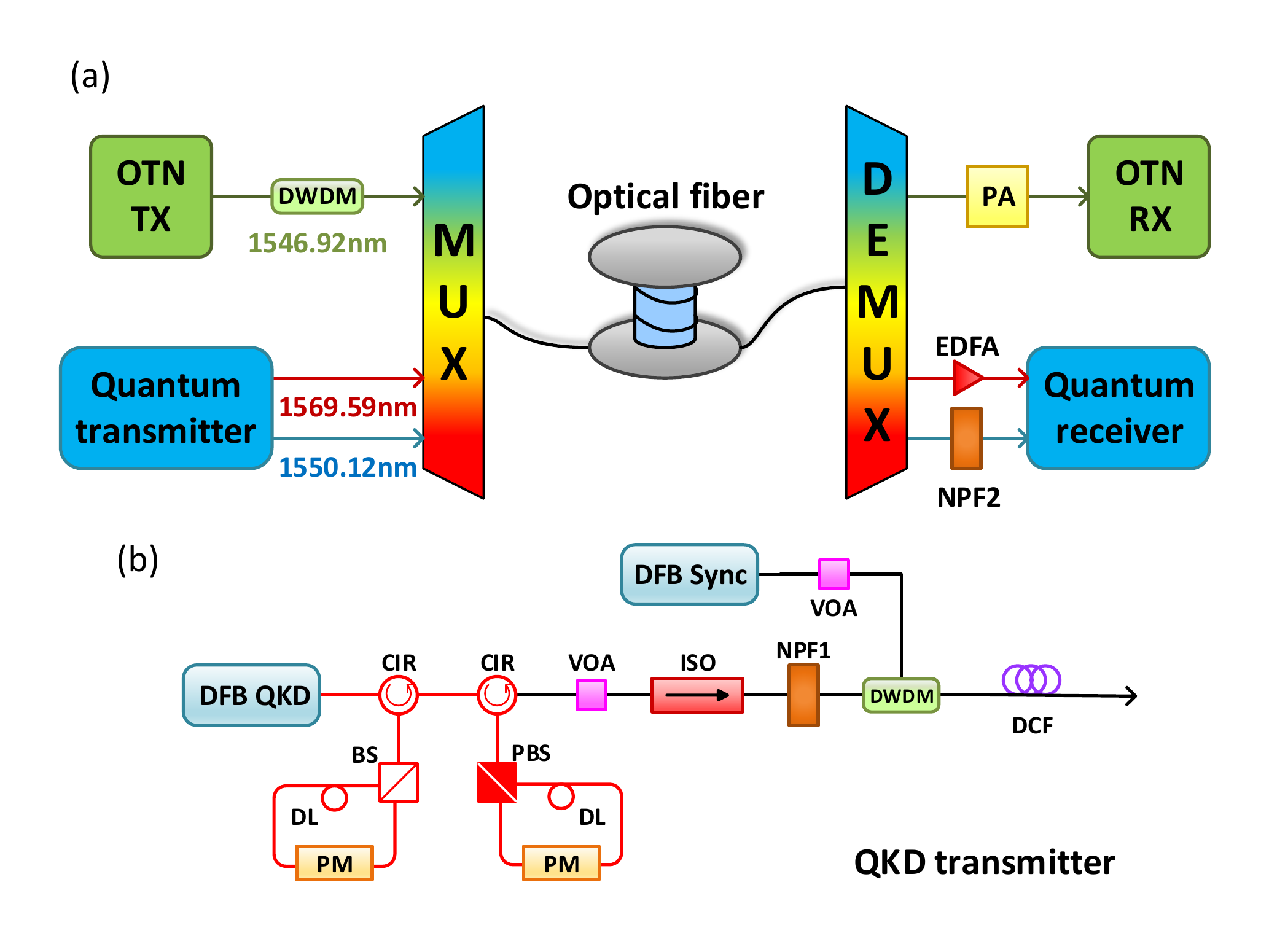}
\caption{Experimental setup. (a) Schematics for multiplexing of QKD and Optical transport network (OTN). TX: transmitter (1546.92 nm), DWDM: dense wavelength division multiplexer, PA: pre-amplifer, RX: receiver, EDFA: erbium-doped fiber amplifier, NPF: narrow passband filter (NPF1: 0.08 nm, NPF2: 0.16 nm). (b) QKD transmitter. DFB: distributed feedback laser, QKD: 1550.12 nm, Sync: synchronization clock (1569.59 nm), CIR: circulator, BS: beam splitter, PBS: polarization beam splitter, DL: delay length, PM: phase modulator, VOA: variable optical attenuator, ISO: isolator, DCF: dispersion compensation fiber. The black line indicates SMF and red line corresponds to polarization-maintaining SMF.}
\end{figure}

Figure 2(a) illustrates the schematic diagram of the experimental setup of QKD coexisting with classical optical communication. The 100 Gbps data channel with a wavelength of 1546.92 nm is provided by a commercial OTN instrument (OPTN8600V). Two DWDMs are employed to filter out the 1550.12 nm wavelength component of the classical data channel. A PA with -33 dBm receiving sensitivity is applied to amplify the light intensity of the classical optical signal to reach the optical power receiving threshold of the OTN-RX.

The QKD system is operated with a repetition rate of 625 MHz based on polarization-encoding decoy-state BB84 protocol\cite{Hwang2003,Wang2005,Lo2005}. In the QKD transmitter, as shown in Fig. 2(b), the intensities of signal state $\mu$, decoy state $\upsilon$, and vaccum state $\omega$ are 0.4, 0.2 and 0 photons/pulse, respectively, with the corresponding emission probability of 6:1:1. Different intensities are adjusted by the first Sagnac interferometer and the four polarization states are produced by the second Sagnac interferometer\cite{Shen2013}. For both Sagnac interferometers, the DL is tuned to 16.3 cm, and optical pulses entering the PM from the short arm of interferometer are modulated. The NPF1 and DCF are employed to reduce the effect of chromatic dispersion on QKD pulses.

The QKD receiver includes polarization beam splitters, polarization controllers and InGaAs/InP single-photon detectors (SPD)\cite{Wang2017}. An EDFA is applied to amplify the Sync pulses to guarantee its detection. The NPF2 with the central wavelength matching the NPF1 is used to filter out SRS noise. Four SPDs are adopted, and the detection efficiency, gate frequency, and dark count rate per gate for each one is 10\%, 1.25 GHz, and $3.0 \times 10^{-7}$, respectively\cite{doi:10.1063/1.4746291,Zhang2015}. Furthermore, stable real-time secure key generation with a block size of 500 kbits is accomplished by post-processing including error correction, error verification, and privacy amplification\cite{Mao:18}.

To integrate QKD and classical signals over FMF, we use the $\rm LP_{01} in$ ($\rm LP_{02} in$) scheme, where the classical data channel and QKD are allocated to the $\rm LP_{01}$ ($\rm LP_{02}$) mode and the $\rm LP_{02}$ ($\rm LP_{01}$) mode, respectively. QKD and Sync pulses are multiplexed into the same LP mode using a 1550.12 nm DWDM. 

As a comparison, the co-propagation of QKD with the same classical data channel is also presented over SMF. The fiber loss of quantum signal and the classical data channel is 0.190 dB/km and 0.192 dB/km, respectively. The MUX/DEMUX for SMF is one 1546.92 nm DWDM with the IL of 0.49/0.36 dB. Considering the difference in IL between the MSC and DWDM, the fiber-input power, which refers to the output power after passing through the MUX, is calibrated to -2.60 dBm. 

\section{Results and discussion}
In the experiment, since there is only one classical data channel allocated to one LP mode with optical power less than 1 mW, the SRS noise is the main factor of limiting the transmission distance of QKD over FMF\cite{AGRAWAL2019621}. As for SMF, Raman noise is also the main limitation for the co-propagation of QKD and classical signals\cite{Frohlich:17}. We measure SRS noise at different distances in three multiplexing schemes, and the average SRS coefficients are calculated as 12076, 2637, and 2655 cps/(mw $\cdot$ km) for SMF, $\rm LP_{01} in$, and $\rm LP_{02} in$, respectively\cite{Chapuran_2009}. Figure 3(a) plots the measured data and the simulated SRS noise as a function of distance.

\begin{figure}[h!]
\centering
\includegraphics[width=\textwidth]{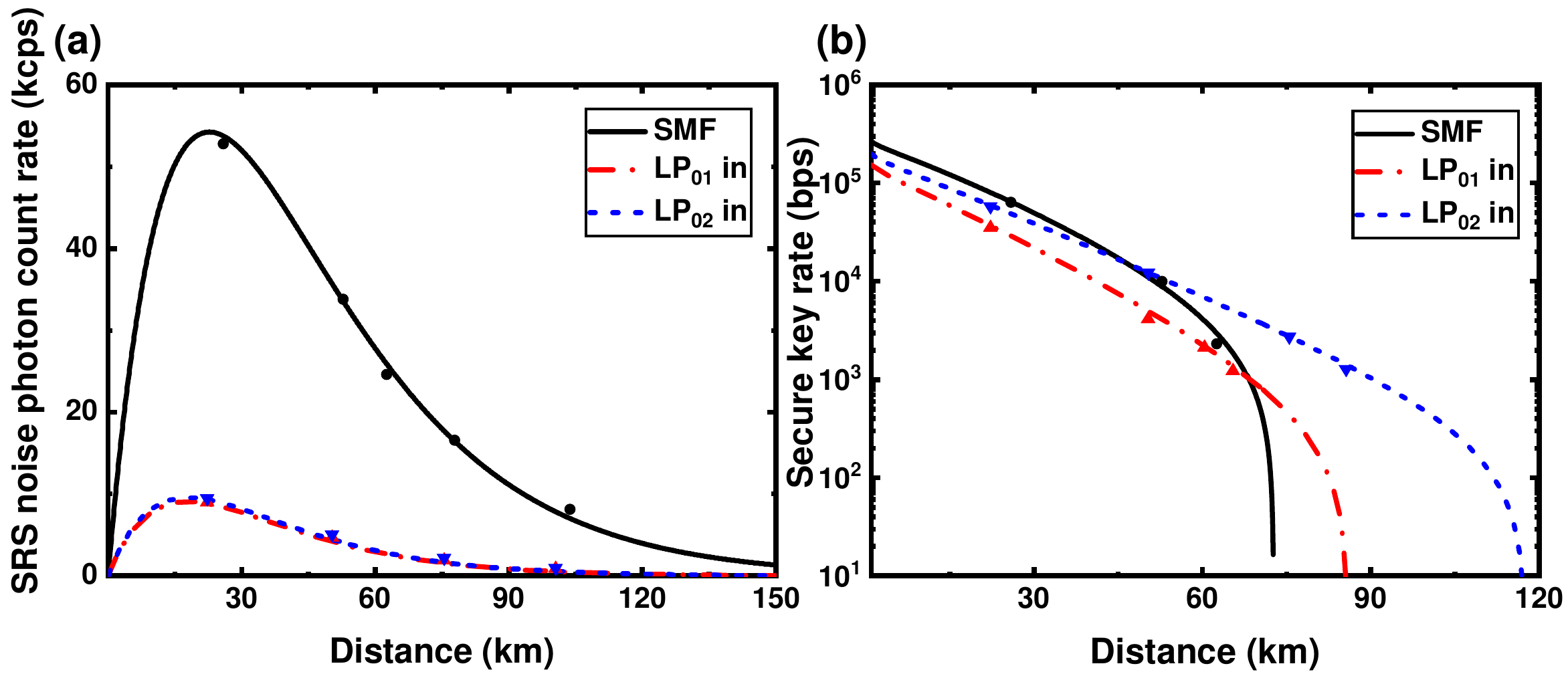}
\caption{SRS noise photon count rate (a), and secure key rate (b) as a function of distance. The black solid, red dash-dotted, and blue dashed line represents the simulation of SMF, $\rm LP_{01} in$, and $\rm LP_{02} in$, respectively. Black solid circles, red solid-upward triangles, and blue solid-downward triangles denote the experimental data for SMF, $\rm LP_{01} in$, and $\rm LP_{02} in$, respectively. }
\end{figure}

Compared with SMF, the SRS noise in FMF is reduced by 86\%. For instance, in the $\rm LP_{02} in$ scheme, the SRS noise generated by the classical signal in the $\rm LP_{02}$ mode is also in the $\rm LP_{02}$ mode with high possibility, which can be effectively filtered after mode DEMUX. In addition, SRS noise can be further suppressed due to the large effective core area of the mode\cite{Carpenter:13}. 

Figure 3(b) plots the measured and simulated  secure key rates over FMF and SMF\cite{Ma2005}. At the secure transmission distance of 63 km, 65 km, and 86 km, we achieve secure key rates of 2.3 kbps, 1.2 kbps, and 1.3 kbps for SMF, $\rm LP_{01} in$, and $\rm LP_{02} in$, with a quantum bit error rate (QBER) of 4.0\%, 3.8\%, and 3.7\%, respectively. At distances less than 45 km,  the secure key rate of SMF is slightly higher than that of FMF, due to the low losses of DWDM and SMF. However,  the QBER caused by excessive SRS noise exceeds the correction efficiency of the system, which results in a sharp drop in the secure key rate over long distance. As for the $\rm LP_{02} in$ scheme, the longest distance transmission is achieved because of the advantage of FMF in suppressing SRS noise. The secure transmission distance of $\rm LP_{01} in$ is restricted by the high loss for QKD pulses assigned to the $\rm LP_{02}$ mode. 

The maximum secure transmission distance of the $\rm LP_{02} in$ scheme can be further extended. Figure 4 plots the simulations of secure key rate, aiming to the possible improvements in the future. The first factor limiting the co-propagation distance is the fiber-input power. The SRS noise generated by classical optical signals is directly proportional to the fiber-input power of classical optical signals\cite{Chapuran_2009}. If the optical power budget of the classical optical communication is sufficient, the fiber-input power can be reduced, which means that the minimum power is the sum of the overall loss of the FMF link and the receiving sensitivity of PA. Particularly, the secure key rate can be increased at short distances, as plotted in the dashed line of Fig. 4.

\begin{figure}[h!]
\centering
\includegraphics[scale=0.4]{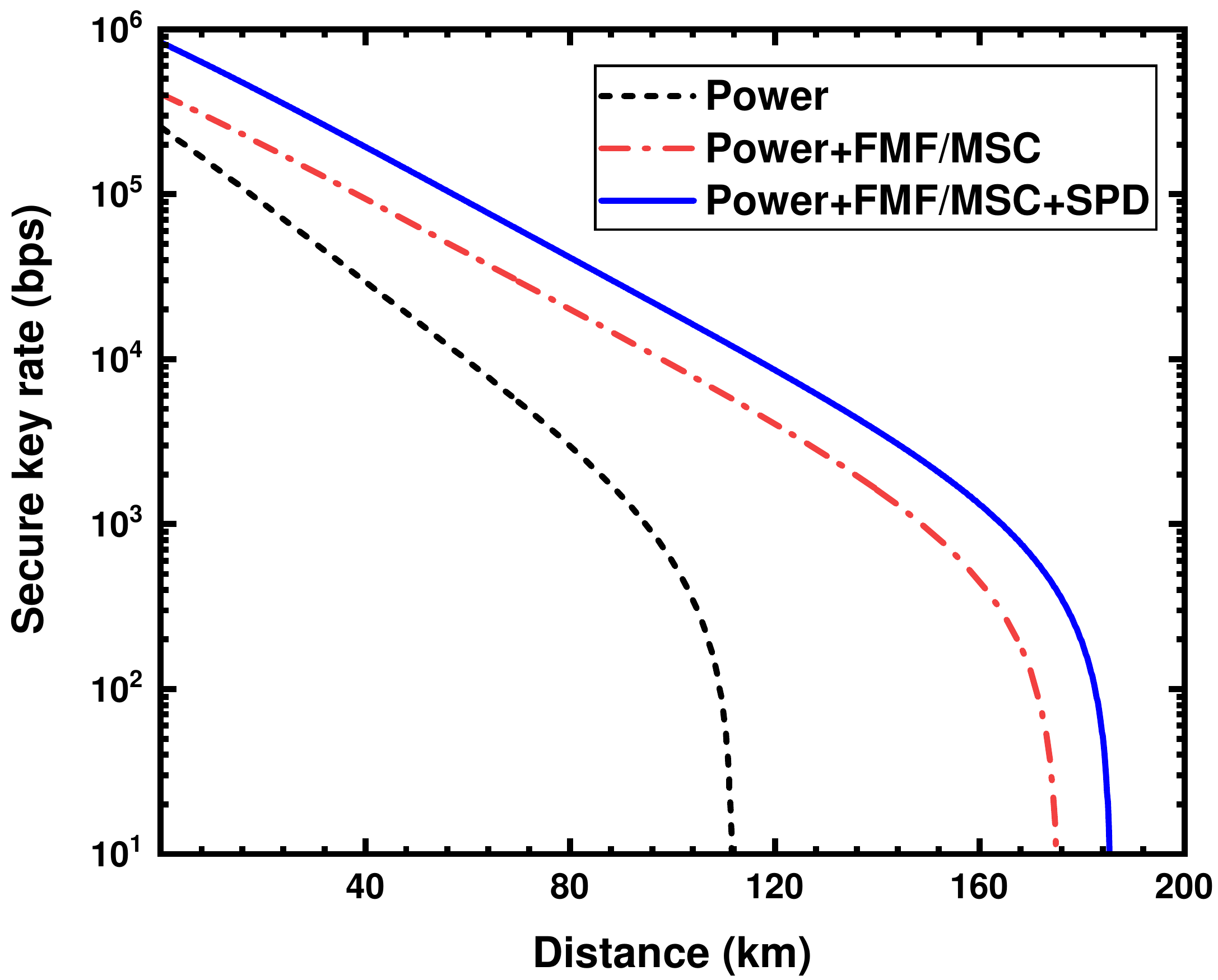}
\caption{Calculated secure key rates of the $\rm LP_{02} in$ scheme with Power (black dashed line), Power+FMF/MSC (red dash-dotted line), and Power+FMF/MSC+SPD (blue solid line) improvements.}
\end{figure}

The second factor limiting the co-propagation distance is the loss of FMF and MSCs. Owing to advances in manufacturing technology, the attenuation coefficient of FMF can in principle reach the level as low as the ultra-low-loss SMF, i.e., 0.16-0.17 dB/km. Similarly, the IL of MSCs can be reduced to the level as the conventional DMDM, i.e., 0.36-0.49 dB. As a result, both the secure key rate and transmission distance are significantly increased, as plotted in the dash-dotted line of Fig. 4.

The third factor limiting the co-propagation distance is the performance of SPD. QKD highly depends on the parameters of SPD. Given the parameters in Ref.\cite{Frohlich:17}, i.e., 20\% detection efficiency and 230 cps dark count rate, as well as the above improvements, the maximum secure transmission distance can reach as long as 185 km, as plotted in the solid line of Fig. 4.

\section{Conclusion}
In summary, we have presented for the first time the QKD implementation coexisting with 100 Gbps data channel over weakly-coupled FMF using the fiber-based MSCs, achieving the transmission distance of 86 km with a secure key rate of 1.3 kbps. We have discussed the possible improvements in the future, and the simulation results show that the maximum secure transmission distance can reach 185 km over FMF. Due to the advantages of FMF, including additional modal isolation, large effective core area, simple fabrication, and cost-effective fiber fusion splicers, our work provides a practical approach to integrate QKD and classical optical communication over the telecommunication fiber-optical infrastructure. 

\section*{Funding}

National Key R $\&$ D Program of China (No. 2017YFA0303903); National Natural Science Foundation of China (Grant No. 61875182).

\section*{Disclosures}
The authors declare that there are no conflicts of interest related to this article.


\end{document}